\documentclass[aps,prd,preprint]{revtex4-1}
\usepackage{graphicx}
\usepackage{amsmath,amssymb,graphics,setspace,array,color}
\usepackage[export]{adjustbox}
\usepackage{commath}
\usepackage[export]{adjustbox}
\usepackage{courier,float}
\usepackage{enumitem}
\usepackage[colorinlistoftodos]{todonotes}
\usepackage[normalem]{ulem}

\newcommand{\mathsym}[1]{{}}
\newcommand{\unicode}[1]{{}}

\makeatletter
\def\p@subsection{}
\def\p@subsubsection{}
\def\p@paragraph{}
\def\p@subparagraph{}
\makeatother

\makeatletter
\def\l@section{\@dottedtocline{1}{1em}{2em}}
\def\l@subsection{\@dottedtocline{2}{1.5 em}{2em}}
\def\l@subsubsection{\@dottedtocline{3}{2em}{3em}}
\def\l@paragraph{\@dottedtocline{4}{2.5em}{4em}}
\def\l@subparagraph{\normalfont \@dottedtocline{5}{3.5 em}{4 em}}
\makeatother

\newenvironment{myitemize}
{ \begin{itemize}
    \setlength{\itemsep}{0pt}
    \setlength{\parskip}{0pt}
    \setlength{\parsep}{0pt}     }
{ \end{itemize}                  } 

\newenvironment{myenumerate}
{ \begin{enumerate}
    \setlength{\itemsep}{0pt}
    \setlength{\parskip}{0pt}
    \setlength{\parsep}{0pt}     }
{ \end{enumerate}                  }

\usepackage{fancyhdr}

\begin{document}

\title{Revisitation of  time delay interferometry combinations that suppress laser noise in LISA}

\author{Martina~Muratore}\affiliation{\addressi}
\author{Daniele~Vetrugno}\affiliation{\addressi}
\author{Stefano~Vitale}\email{contact: stefano.vitale@unitn.it}\affiliation{\addressi}

\def\addressi{Dipartimento di Fisica, Universit\`a di Trento and Trento Institute for 
Fundamental Physics and Application / INFN, 38123 Povo, Trento, Italy}
\date{May 13, 2020}
\begin{abstract}
With the purpose of understanding how time delay interferometry (TDI)  combinations can  best be used for the characterisation of LISA instrumental noise,  we revisit their laser frequency  noise cancellation properties. 
We have developed an algorithm to search for all possible combinations that suppress noise at the same level as the X, Y and Z classical combination. The algorithm calculates delays using symbolic formulas that explicitly include  velocities and accelerations of satellites up to the relevant order.  In addition, once a combination has been identified, delays are verified by solving numerically the relevant equations  using Keplerian orbits and Shapiro delay corrections. We find that the number of combinations that suppress the noise is larger than what was reported in the literature. In particular we find that some  combinations that were thought to only partly suppress the noise, in reality do suppress it at the same level of accuracy as the basic X, Y and Z combinations.    
\end{abstract}

\maketitle

\section{Introduction}

With the aim of participating to the effort for  the preparation for LISA \cite{LISA2017} data analysis, we are studying  methods to calibrate the noise during operations, and to discriminate  spurious signals from gravitational waves.

One key element in the LISA  data production chain is the post processing technique called Time Delay Interferometry (TDI)\cite{Armstrong_1999} aimed at suppressing the intense laser frequency noise. Data from the 6 independent laser links connecting the three satellites \cite{LISA2017}, are properly time shifted and combined  to form the  final GW signal. This post-processing technique  circumvents the impossibility of physically building in space an equal arm Michelson interferometer, which  would intrinsically beat the frequency noise by comparing light generated at the same time.

Many possible TDI combinations that suppress  the frequency noise have been identified in the past \cite{Shaddock2003,Vallisneri}. Those combinations have different sensitivities to GW signals. Some are indeed rather insensitive and promise to be useful for the characterisation of instrument noise.

As instrument noise is expected to have multiple, independent sources, we are strongly interested to exploit all possible combinations that could allow to discriminate among those sources, and between them and the GW signal. We  have thus revisited all possible TDI configurations that suppress the laser noise contribution to the level required by LISA. 

To this purpose, we have  developed our own search and verification algorithm, and used  it to  explore the space of possible configurations and to find those that fulfil the noise suppression requirements.

This algorithm finds  a certain number of 12 link TDI combinations that fulfil the noise suppression requirement as accurately as  standard 16 link  combinations, like, for instance, those known as X, Y and Z \cite{Shaddock2003}.  This is at variance with  what had been reported in the literature \cite{Shaddock2003,Vallisneri}, where 12 link combinations were found  to suppress the noise to a lesser extent than the  16 link combinations. In addition we also find a larger number of 16-link combinations that suppress noise at comparable level than those reported in the literature.

This note describes our approach, reports the results,  and discusses the source of this apparent discrepancy.

\section{\label{STDI} Noise suppressing TDI combinations}

In order to assess if a given TDI combination suppresses the laser frequency noise, we adopt a home-made version of the method of ref. \cite{Vallisneri}. 
The basis for a  TDI combination that suppresses the laser noise is, in our method, a set of space-time events placed on the worldlines of the three satellites. 

\begin{figure}[h]
\begin{center}
\includegraphics[width=.8\textwidth]{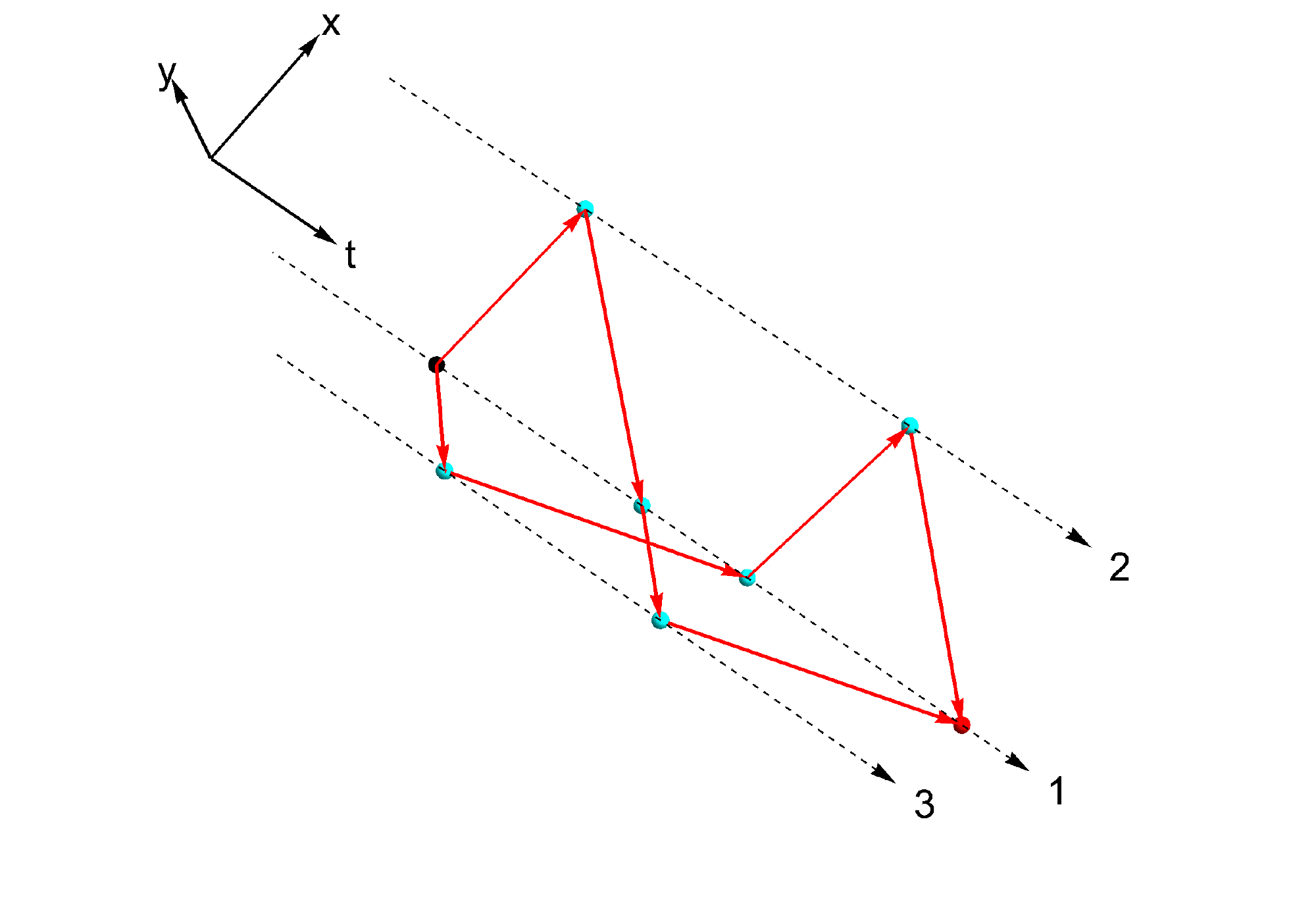}
\end{center}
\caption{Space-time diagram of a TDI combination. The dashed lines represent the worldlines of the satellites. The  dots represent the events. Different colours represent events of different nature as explained in the text. The red arrows represent LISA links, that is, light beams propagating between two events. The space-time length of each arrow is zero. The specific picture would represent the first generation X  combination for stationary satellites.}
\label{Figarr}
\end{figure}

Each event on a specific worldline  must have a null space-time separation from other two events belonging to the  same set, but placed on different wordlines. This way all events may be connected by LISA laser links, each event being at one of the ends of two different links (See fig. \ref{Figarr}). 
The events falls in  the three categories listed below.
\begin{myenumerate} 
\item{The event may represent the simultaneous emission of two beams from one satellite toward the other two. This is the case of the event marked by the black dot in fig. \ref{Figarr}, where both links originate from the event.}
\item{\label{a1} The event may represent the simultaneous reception, at one satellite,  of two beams originating at two different satellites. This is the case of the red dot in fig. \ref{Figarr}, where both links terminate at that event.  Such an event represents a measurement of the  difference of phase between the two received beams.}
\item{One link terminates at the event and one  originates from it, as is the case for the cyan dots of fig. \ref{Figarr}. This occurs when a satellite receives a beam from another satellite, measures the difference of phase of such received beam to the local laser, and then beams the local laser, either back to that same satellite,  like in a transponder, or over to the third satellite, like in a transmission relay.}
\end{myenumerate}

A TDI sequence may contain more than one pair of emission/measurement events. The final TDI output is the sum of all the  measurements performed at the measurement events. 

One can check that the phase noise of the laser at any event, whatever its category, enters twice in the final TDI output, but with opposite signs. As a consequence its contribution to the output cancels out. \footnote{We are obviously ignoring the relative frequency noise between the two lasers, placed on the same satellite, but  belonging to  two different arms, the so called back-link noise. 
However the contribution of this  noise source is not affected by the choice of delays, and does not enter then in the evaluation of the effectiveness of a TDI combination.}

As clearly shown by  fig. \ref{Figarr}, one way of looking at a TDI combination, is that the links form a closed path in space-time that one can follow  starting from any of the events, passing sequentially by all the remaining events in the set, and eventually getting back to the initial one \footnote{We are excluding the possibility that a smaller subset of  events constitutes a closed path. In such case, the subset would form a proper TDI combination by itself}. Along such a path, some links are followed from emission to reception, but some must be followed in the reverse order.  Thus, along the path, the events  are not ordered in time, otherwise the path could not both start and end at the same event, and then at the same time.

\begin{figure}[h]
\begin{center}
\includegraphics[width=1\textwidth]{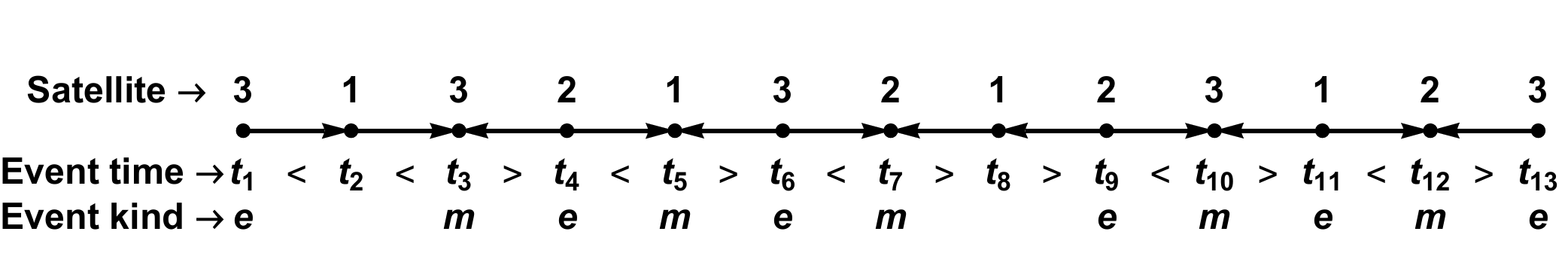}
\end{center}
\caption{Schematic of a possible TDI sequence. Arrows  indicate links and points indicate events. Arrows directions indicate those of the corresponding links. Label `e' stands for emission and label `m' stands for measurement. Unmarked events are of transponder/relay kind. The figure also indicates the relative time ordering of two contiguous events. If the sequence is a proper TDI sequence, then events 1 and 13 happen on the same satellite and $t_1=t_{13}$. The 12 link sequence in the picture turns out to be a proper TDI sequence.}
\label{Figseq}
\end{figure}

Thus, to search for  a proper  TDI combination, one may lay down  a  sequence of N events, connected by N-1 links, such that  two contiguous  events are on different satellites and have  null space-time separation.  Link number $k$ may originate at event $k$ and terminate at event $k+1$, in which case $t_k < t_{k+1}$, or viceversa, and then  $t_k >t_{k+1}$. If the sequence ends at the same event from which it starts, then it represents a proper TDI combination (See fig. \ref{Figseq}). 

In reality the sequence will always end  at an event that is on the same satellite as the starting one, but at a slightly  different time.  In order to suppress the laser noise to the required accuracy, such time difference must be kept below $\simeq 3 ns$, i.e. one light-meter. Actually, the current best estimate for the overall timing precision that LISA should be able to achieve is $\simeq 0.3 ns$ \cite{LISAPDD}. This figure results from a combination of  inter-satellite ranging error, of on-board  anti-aliasing filtering, and of on ground re-interpolation of Hz-sampled data \cite{filters}.
\section{\label{Search} The search algorithm and the calculation of time delays}

In order to implement the programme above, one needs to estimate the  light  propagation  delay along each link within the candidate TDI sequence. 
These delays  depend on time, as both the distances among the satellites, and their distances to the Sun, are not constant. In the literature \cite{Shaddock2003,Vallisneri} it is assumed that the time of propagation $\Delta t(t)$ of a beam emitted by satellite $i$, and received at time $t$ at  satellite $j$,   can be approximated with sufficient accuracy as:
\begin{equation}
\Delta t(t)=\frac{1}{c}\left(L_{ij}+\dot{L}_{ij}t\right),
\label{Edl}
\end{equation}
during the entire time it takes to the light to propagate throughout a TDI combination.

One can then propagate eq. \ref{Edl} along the full closed path representing a TDI combination, neglect at each step terms of order $\left(\dot{L}/c\right)^2$ or higher, and test if the sum of the delays is zero.

This can be done symbolically, i.e. without the need of assigning  specific numerical values to $L_{ij}$ and $\dot{L}_{ij}$, which makes the algorithm very efficient.
This is the idea at the basis of the algorithm used, for instance, in  \cite{Vallisneri}.

\subsection{\label{Delay} Time delays  formulas in different reference frames}

In the spirit of carefully verifying all the steps of such a procedure, we decided to work out the explicit dependence of $L_{ij}$ and  $\dot{L}_{ij}$ on positions, velocities and accelerations of  satellites, and expand them to the proper order.

In the standard heliocentric (i.e. celestial isotropic) reference frame, to  null the space time separation between the event of beam emission at time $t_e$ with space coordinates $\vec{r}_e(t_e)$, and that of beam reception at time $t_r$ and space coordinates $\vec{r}_r(t_r)$,    it is sufficient to solve the equation:
\begin{equation}
\begin{split}
\vert \vec{r}_r\left(t_r\right)-\vec{r}_e\left(t_e\right)\vert+R_\odot Log\left(\frac{\vert \vec{r}_r\left(t_r\right)\vert+\vert \vec{r}_e\left(t_e\right) \vert +\vert \vec{r}_r\left(t_r\right)-\vec{r}_e\left(t_e\right) \vert}{\vert \vec{r}_r\left(t_r\right)\vert+\vert \vec{r}_e\left(t_e\right) \vert -\vert \vec{r}_r\left(t_r\right)-\vec{r}_e\left(t_e\right) \vert}\right)= c \left(t_r-t_e\right),
\end{split}
\label{Enull}
\end{equation}
where all the  relativistic corrections have been  adsorbed in the  Shapiro delay formula, which is the term preceded by  the Sun Schwarzschild radius  $R_\odot$  \cite{Moyer}.

The expansion of  eq. \ref{Enull} to the relevant order produces a rather cumbersome formula, which is explicitly reported in eq. \ref{Elldot} on page \pageref{Elldot} in the appendix. Such a formula is impractical for the symbolic search algorithm we need here.

The calculation of delays, however, can be done in any coordinate frame, as the simultaneity between two events,  occurring at the same satellite, remains true in any frame.

It is important to stress  though that  this choice of reference frame has nothing to do with the practical implementation of TDI. That implementation will be based on data  time-stamping and pseudo-ranging measurements that all happen on board individual satellites, and will not require to pick any global frame to achieve noise cancellation. The choice of the frame is required here, and in the literature, only to verify the existence of TDI combinations that suppress the noise.

We show in the appendix that formulas simplify considerably  in the  Fermi normal frame attached to a point particle orbiting the Sun \cite{Ashby}, in the neighbourhood of the center of mass (COM) of the LISA constellation. 

In  such \emph{quasi-inertial} system there are no  inertial effects,  because  its axes are still non rotating relative to the celestial frame, but the gravity of the Sun is cancelled to first order as the coordinate origin is in free-fall. 

The residual deviation  from a Minkowski metric  is either of order $\left(R_\odot /R\right)\left( d^2/R^2\right)$ or  smaller. Here $R$ is the distance of the COM from the Sun and $d$ is the size of the constellation. This is a suppression of non-flatness, and then of Shapiro delay, by a factor $\simeq10^{-4}$ relative to the isotropic celestial frame.

Notice that in such a frame the time coordinate is the proper time of the particle orbiting the Sun, and  differs from the celestial time coordinate by a scale factor \cite{Ashby} which is irrelevant here. 

In addition velocities and accelerations are significantly suppressed, leaving a very simple formula for the total delay.
Such formula is calculated in the appendix  sect. \ref{A1}, and  is: 
\begin{equation}
\begin{split}
&\Delta t\left(t\right)=\frac{\vert \Delta \vec{r}\vert}{c}\times\left(1+\frac{\vec{v}}{c}\cdot \Delta\hat{r}\right)+\Delta\vec{v}\cdot\Delta\hat{r}\; t
\label{Efinbis}
\end{split}
\end{equation}
Here, as  shown in eq. \ref{Epos2} in the appendix,    $\Delta \vec{r}$ is the space vector separating the satellites at time $t=0$, $\Delta \vec{v}$ is their relative velocity at the same time, $\vec{v}$ is the velocity of the emitter satellite, again at $t=0$, and, finally,  $\Delta\hat{r}$  is the unit vector parallel to $\Delta \vec{r}$.

Translating to  the language of eq. \ref{Edl}, eq. \ref{Efinbis} converts into:
\begin{equation}
\begin{split}
&L_{ij}=\vert \Delta \vec{r}_{ij}\vert\times\left(1+\frac{\vec{v}_{i}}{c}\cdot \Delta\hat{r}_{ij}\right)\\
&\dot{L}_{ij}=\Delta\vec{v}_{ij}\cdot\Delta\hat{r}_{ij}
\label{sym}
\end{split}
\end{equation}
The formula above has allowed us to implement a simple and practical symbolic algorithm to search for TDI combinations.

\subsection{\label{Numeric} Numerical verification of the goodness of a TDI combination}
 
As the symbolic formula contains many approximations,  we decided to verify numerically the results of the symbolic algorithm, by calculating the total delay of each TDI combinations on the actual satellite orbits.
To do that we check that the coordinates of the satellites on their actual orbits, at the times of the various events in the sequence, fulfil the identity in eq. \ref{Enull} to the right numerical accuracy.
This calculation is done in the celestial isotropic frame using the classical LISA orbits \cite{Hughes}, with a mean arm-length of $L=2.5\times10^6 \text{km}$.

The calculation starts from an  event at one end of one of the links in the sequence. The event happens at a time $t_e$ that can be arbitrarily selected.
The event  space coordinates $\vec{r}_e(t_e)$ are derived from the orbit equation for the specific satellite on which the event takes place. 

One looks then for the closest time $t_r$ at which the coordinates  $\vec{r}_r(t_r)$ of the satellite at the other end of the link, fulfil, together with $\vec{r}_e(t_e)$ and $t_e$,  eq. \ref{Enull}. $t_r$ and $\vec{r}_r(t_r)$ become then the  coordinates of  the next event in the sequence. Equivalently, if the link is followed in a reverse order, one starts from $t_r$ and   $\vec{r}_r(t_r)$, and calculates  $\vec{r}_e(t_e)$ and $t_e$.

This procedure is iterated along the entire sequence of links and events, until the starting satellite is reached again. In order for the sequence to be a proper one, the time difference between this final event and the starting one must be less than the required minimum delay.

As one is interested to an accuracy of nano-seconds  over a total propagation time of about hundred seconds, it is important to keep the precision of the calculation under control. With this we mean that the  two sides of eq. \ref{Enull} must be calculated with enough digits, $\gtrsim 13$ that numerical rounding does not affect the results, even after iterating the calculation over the entire TDI sequence. To be on the safe side we do calculations with femto-second  precision.  

Accuracy is less important. Indeed, orbits locally approximated by  simple second order time polynomials give the same results  as  the `exact' LISA orbits, i.e. the same TDI combinations passes the test for both choices of the orbits.  This provided that satellites separations, velocities and acceleration are of the same order as those in the `exact' LISA orbits. This is consistent with the observation that the problem is locally well described by eq. \ref{Epos2} in the appendix.

We must say though that we kept also the accuracy of the orbit to a comparable level, expanding the solution of the Kepler equation up to the sixth order in the eccentricity. This is definitely more accurate than what we need, but gives us a good margin at the expense of a negligible calculation load. 

In summary we have calculated delays with a precision better than  pico-seconds, while accuracy is probably of order  of a  pico-second, limited by all approximation contained in eq. \ref{Enull}.

All the TDI combinations that have been found by the symbolic search algorithm, have also  passed the numerical verification test.

\section{\label{results}Results}

Our search algorithm finds many TDI combinations for which total delay is of order of pico-seconds or less. We find a large number of combinations that include 16 links, but we also find combinations that only include 12 links.

\begin{figure}[h]
\includegraphics[width=0.8\textwidth]{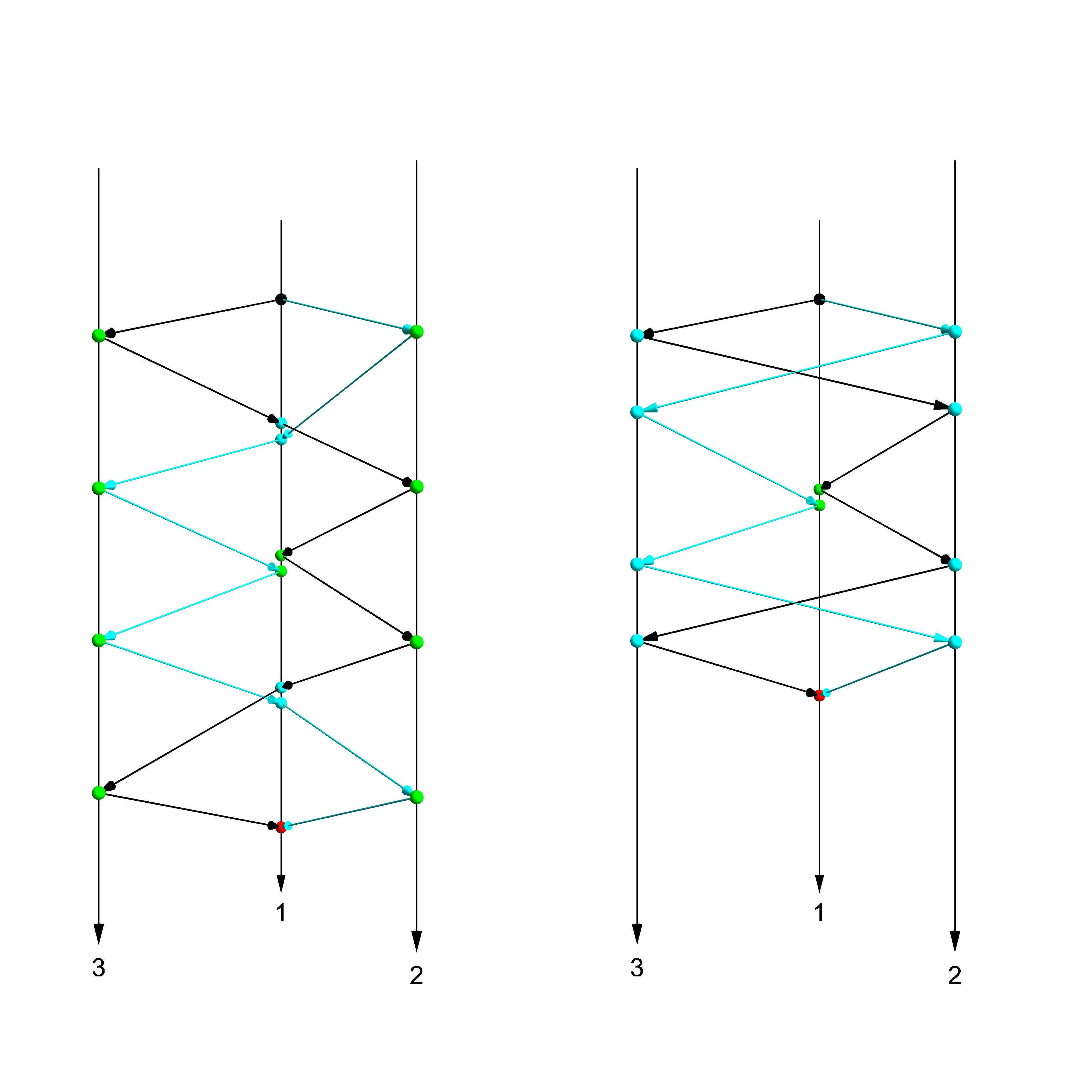}
\caption{ Space-time schematics for the second generation X combination (left) and for the $\alpha$ combination (right)\cite{Shaddock2003}. The black, numbered lines represent the satellite world-lines.The coloured points represent the events. Black: emission; green: relay; cyan: transponder; red: measurement. Time intervals are not to any scale.}
\label{FigX}
\end{figure}

As an example, the  outputs of our search algorithm for the standard, second generation X combination, and  for the 12-link $\alpha$ combination, as defined in \cite{Shaddock2003}, are illustrated in  table \ref{T:X} and in fig. \ref{FigX}. 

\begin{table}[h]
\begin{center}
\resizebox*{!}{.6\textheight}{
\begin{tabular}{|c|c|c|c|c||c|c|c|c|c|}
\hline
\multicolumn{5} {|c|} {X}&\multicolumn{5} {|c|} {$\alpha$}\\
\hline
 Event&Satellite&Time&Event&Link&Event&Satellite&Time&Event&Link\\
  number&&[s]&kind&direction& number&&[s]&kind&direction\\
\hline
\hline
 1 & 1 & 0 & e &  & 1 & 1 & 0 & e &  \\ \hline
  &  &  &  & $\downarrow$  &  &  &  &  & $\downarrow$  \\ \hline
 2 & 3 & 8.32 & t &  & 2 & 3 & 8.32 & r &  \\ \hline
  &  &  &  & $\downarrow$  &  &  &  &  & $\downarrow$  \\\hline
 3 & 1 & 16.6 & r &  & 3 & 2 & 16.7 & r &  \\\hline
  &  &  &  & $\downarrow$  &  &  &  &  & $\downarrow$  \\\hline	
 4 & 2 & 25.0 & t &  & 4 & 1 & 25.0 & t &  \\\hline
  &  &  &  & $\downarrow$  &  &  &  &  & $\downarrow$  \\\hline
 5 & 1 & 33.3 & t &  & 5 & 2 & 33.4 & r &  \\\hline
  &  &  &  & $\downarrow$  &  &  &  &  & $\downarrow$  \\\hline
 6 & 2 & 41.7 & t &  & 6 & 3 & 41.7 & r &  \\\hline
  &  &  &  & $\downarrow$  &  &  &  &  & $\downarrow$  \\\hline
 7 & 1 & 50.0 & r &  & 7 & 1 & 50.0 & m &  \\\hline
  &  &  &  & $\downarrow$  &  &  &  &  & $\uparrow$  \\\hline
 8 & 3 & 58.3 & t &  & 8 & 2 & 41.7 & r &  \\\hline
  &  &  &  & $\downarrow$  &  &  &  &  & $\uparrow$  \\\hline
 9 & 1 & 66.7 & m &  & 9 & 3 & 33.3 & r &  \\\hline
  &  &  &  & $\uparrow$  &  &  &  &  & $\uparrow$  \\\hline
 10 & 2 & 58.3 & t &  & 10 & 1 & 25.0 & t &  \\\hline
  &  &  &  & $\uparrow$  &  &  &  &  & $\uparrow$  \\\hline
 11 & 1 & 50.0 & r &  & 11 & 3 & 16.7 & r &  \\\hline
  &  &  &  & $\uparrow$  &  &  &  &  & $\uparrow$  \\\hline
 12 & 3 & 41.6 & t &  & 12 & 2 & 8.35 & r &  \\\hline
  &  &  &  & $\uparrow$  &  &  &  &  & $\uparrow$  \\\hline
 13 & 1 & 33.3 & t &  & 13 & 1 &$4.18\times10^{-13}$ & e &  \\\hline
  &  &  &  & $\uparrow$  &  &  &  &  &  \\\hline
 14 & 3 & 25.0 & t &  &  &  &  &  &  \\\hline
  &  &  &  & $\uparrow$  &  &  &  &  &  \\\hline
 15 & 1 & 16.7 & r &  &  &  &  &  &  \\\hline
  &  &  &  & $\uparrow$  &  &  &  &  &  \\\hline
 16 & 2 & 8.35 & t &  &  &  &  &  &  \\\hline
  &  &  &  & $\uparrow$  &  &  &  &  &  \\\hline
 17 & 1 & $1.25\times10^{-12}$ & e &  &  &  &  &  &  \\\hline
\end{tabular}
}
\end{center}
 \caption{Output of the search and verification algorithm for the second generation X combination (left) and for the $\alpha$ combination (right) as defined in \cite{Shaddock2003}. The table lists the events that compose the sequence, giving the satellite they take place on, their time of occurrence, and the type of event (e: emission; r: relay; t: transponder; m: measurement). Arrows between the rows give the direction of the link connecting the two events. Calculation correspond to day 123 of the orbit.}
 \label{T:X}
\end{table}

The table shows that total delay  can be suppressed to  same pico-second level both for X and $\alpha$, this last result being at variance with the literature \cite{Shaddock2003,Vallisneri}.

To check that these results do not depend significantly on the selected day, we have repeated the calculation for ten more days, randomly picked along the LISA orbit.
As an example, the time difference between the first and the last event, for both the second generation X and for $\alpha$, are reported in table \ref{tabledelta} . In the same table, we report, for comparison, the sum of delays calculated for a first generation X, and even for the standard Michelson one would do for a LISA with fixed and equal arm-length. As expected, the residuals for those combination coincide, in order of magnitude, with the values, calculated from the orbits, of 
$\left(\vec{v}_r -\vec{v}_e \right)\cdot\left(\vec{r}_r\left(t_r\right)-\vec{r}_e\left(t_e\right)\right)/c^2$ and $\Delta\left\vert\vec{r}_r\left(t_r\right)-\vec{r}_e\left(t_e\right)\right\vert$ respectively. Here $\Delta\left\vert\vec{r}_r\left(t_r\right)-\vec{r}_e\left(t_e\right)\right\vert$ is the difference of the length of the two arms used for the Michelson.  
\begin{table}[h]
\begin{center}
\begin{tabular}{|c|c|c|c|c|}
\hline 
Day & $\Delta T_\alpha [ps] $ &$ \Delta T_{X} [ps]$& $ \Delta T_{X1} [\mu s]$ & $ \Delta T_{X0} [s]$ \\ \hline
9 & -0.32 & 0.34 & -0.21 & 0.010 \\ \hline 
103 & -0.64 & 1.6 & 0.042 & 0.064 \\ \hline
124 & -0.78 & 1.1 & 0.12 & 0.055 \\ \hline 
166 & -0.40 & -0.031 & 0.21 & 0.018 \\ \hline 
171 & -0.17 & 0.14 & 0.21 & 0.013 \\ \hline
198 & -0.15 & 0.096 & 0.21 & -0.017 \\ \hline
 249 & 0.43 & -2.4 & 0.089 & -0.060 \\ \hline 
 302 & 0.82 & -2.4 & -0.10 & -0.058 \\ \hline
 317 & -0.45 & -0.39 & -0.15 & -0.048 \\ \hline
 335 & 1.4 & -0.68 & -0.19 & -0.032 \\ \hline 
\end{tabular}
\end{center}
\caption{Residual delay mismatch $\Delta T_\#$ for TDI combination $\#$, at random selected days during one year of LISA orbits. Besides the values for $\alpha$ and $X$, we also plot the results for the first generation X, $X1$, and for a standard Michelson, $X0$.}
\label{tabledelta}
\end{table}

We also find  other 12 link combinations, besides $\alpha$, that cancel noise  within the same accuracy as $\alpha$. Once the set of all noise canceling 12-link combinations has been purged from obvious internal symmetries, like a circular shift in the sequence, we remain with  12 such combinations in total. By reducing the set also for satellite permutations, 3 combinations remain, all the others being obtained from these 3 by a permutation of the three satellites, possibly followed by a circular shift of the sequence.

One of these three combinations is $\alpha$. The remaining combinations differ, one from the other and from $\alpha$, by the number of pairs of emission-measurement events they include. One  includes two such pairs, while the other includes five  of them. 
These sequences  are illustrated in table \ref{T:12l}.
\begin{table}[h]
\begin{center}
\resizebox{1\textwidth}{!}{
\begin{tabular}{|c|c|c|c|c||c|c|c|c|c|}
\hline
 Event&Satellite&Time&Event&Link&Event&Satellite&Time&Event&Link\\
  number&&[s]&kind&direction& number&&[s]&kind&direction\\
\hline
\hline
  1 & 1 & 0 & e &  & 1 & 1 & 0 & e &  \\\hline
  &  &  &  & $\downarrow$  &  &  &  &  & $\downarrow$  \\\hline
 2 & 2 & 8.35 & r &  & 2 & 2 & 8.35 & m &  \\\hline
  &  &  &  & $\downarrow$  &  &  &  &  & $\uparrow$  \\\hline
 3 & 3 & 16.7 & r &  & 3 & 3 & -0.00158 & e &  \\\hline
  &  &  &  & $\downarrow$  &  &  &  &  & $\downarrow$  \\\hline
 4 & 1 & 25.0 & m &  & 4 & 1 & 8.32 & m &  \\\hline
  &  &  &  & $\uparrow$  &  &  &  &  & $\uparrow$  \\\hline
 5 & 2 & 16.7 & r &  & 5 & 2 & -0.0291 & e &  \\\hline
  &  &  &  & $\uparrow$  &  &  &  &  & $\downarrow$  \\\hline
 6 & 3 & 8.32 & t &  & 6 & 3 & 8.32 & t &  \\\hline
  &  &  &  & $\uparrow$  &  &  &  &  & $\downarrow$  \\\hline
 7 & 2 & -0.0291 & r &  & 7 & 2 & 16.7 & m &  \\\hline
  &  &  &  & $\uparrow$  &  &  &  &  & $\uparrow$  \\\hline
 8 & 1 & -8.38 & e &  & 8 & 1 & 8.32 & e &  \\\hline
  &  &  &  & $\downarrow$  &  &  &  &  & $\downarrow$  \\\hline
 9 & 3 & -0.0576 & r &  & 9 & 3 & 16.6 & m &  \\\hline
  &  &  &  & $\downarrow$  &  &  &  &  & $\uparrow$  \\\hline
 10 & 2 & 8.29 & r &  & 10 & 2 & 8.29 & e &  \\\hline
  &  &  &  & $\downarrow$  &  &  &  &  & $\downarrow$  \\\hline
 11 & 1 & 16.6 & m &  & 11 & 1 & 16.6 & m &  \\\hline
  &  &  &  & $\uparrow$  &  &  &  &  & $\uparrow$  \\\hline
 12 & 3 & 8.32 & t &  & 12 & 3 & 8.32 & t &  \\\hline
  &  &  &  & $\uparrow$  &  &  &  &  & $\uparrow$  \\\hline
 13 & 1 & $-4.98\times10^{-14}$ & e &  & 13 & 1 & $-2.24\times10^{-13} $& e &  \\\hline
\end{tabular}
}
\end{center}
 \caption{Left:  TDI sequence with two pairs of emission/measurement events. Right: TDI sequence with five pairs of emission/measurement events. Definitions are the same as those for table \ref{T:X}}
 \label{T:12l}
\end{table}

Our algorithm finds 174 independent 16 link combinations, once the set has been purged from  internal symmetries. If one also purges the set for all possible permutations among the three satellites, only 35 combinations survive. 
The combinations within this minimum set   fall in a few simple categories, marked by the number of inter-satellite links $N_l$ (out of the six possible ones $1\to2,\;2\to1,\;1\to3,\;3\to1,\;2\to3,\;\text{and}\;3\to2$) that are involved in the sequence, 4 or 6, and by the number of emission/measurements event pairs $N_m$, which ranges from 1 to 6. For instance the X combination only involves 4 links, as the third arm does not enter into it, and 1 final measurement event. The list of all 35 combinations is shown in table \ref{T16l}, with X, the  combination of minimum $N_l+N_m$, showing up on the first line.
\begin{table}[h]
\begin{center}
\resizebox{\textwidth}{.2\textheight}{%
\begin{tabular}{|c|c|c|c|c|c|c|c|}
\hline
$\text{N}_s$&Sequence&$\text{N}_l$&$\text{N}_m$&$\text{N}_s$&Sequence&$\text{N}_l$&$\text{N}_m$\\ \hline
\hline
 1 & \text{1$\rightarrow $3$\rightarrow $1$\rightarrow $2$\rightarrow $1$\rightarrow $2$\rightarrow $1$\rightarrow $3$\rightarrow $1$\leftarrow $2$\leftarrow $1$\leftarrow $3$\leftarrow $1$\leftarrow $3$\leftarrow $1$\leftarrow $2$\leftarrow $1} & 4 & 1 & 19 & \text{1$\rightarrow $3$\rightarrow $1$\leftarrow $2$\leftarrow $1$\leftarrow $3$\rightarrow $2$\rightarrow $1$\rightarrow $2$\leftarrow $3$\leftarrow $2$\rightarrow $1$\rightarrow $2$\rightarrow $3$\leftarrow $1$\leftarrow $2$\leftarrow $1} & 6 & 3 \\ \hline
 2 & \text{1$\rightarrow $3$\rightarrow $1$\leftarrow $2$\leftarrow $1$\leftarrow $3$\leftarrow $1$\rightarrow $2$\rightarrow $1$\rightarrow $3$\rightarrow $1$\rightarrow $2$\rightarrow $1$\leftarrow $3$\leftarrow $1$\leftarrow $2$\leftarrow $1} & 4 & 2 & 20 & \text{1$\rightarrow $3$\rightarrow $1$\leftarrow $2$\rightarrow $3$\rightarrow $1$\rightarrow $2$\leftarrow $3$\leftarrow $1$\leftarrow $3$\leftarrow $1$\leftarrow $3$\leftarrow $2$\rightarrow $1$\rightarrow $3$\rightarrow $2$\leftarrow $1} & 6 & 3 \\ \hline
 3 & \text{1$\rightarrow $3$\rightarrow $1$\rightarrow $3$\rightarrow $1$\leftarrow $2$\leftarrow $1$\leftarrow $3$\leftarrow $1$\rightarrow $2$\rightarrow $1$\rightarrow $2$\rightarrow $1$\leftarrow $3$\leftarrow $1$\leftarrow $2$\leftarrow $1} & 4 & 2 & 21 & \text{1$\rightarrow $3$\rightarrow $1$\rightarrow $3$\rightarrow $1$\leftarrow $2$\leftarrow $3$\leftarrow $1$\rightarrow $2$\rightarrow $3$\leftarrow $1$\leftarrow $3$\rightarrow $2$\rightarrow $1$\leftarrow $3$\leftarrow $2$\leftarrow $1} & 6 & 3 \\ \hline
 4 & \text{1$\rightarrow $3$\rightarrow $1$\rightarrow $2$\rightarrow $3$\rightarrow $2$\rightarrow $1$\rightarrow $3$\rightarrow $1$\leftarrow $2$\leftarrow $3$\leftarrow $1$\leftarrow $3$\leftarrow $1$\leftarrow $3$\leftarrow $2$\leftarrow $1} & 6 & 1 & 22 & \text{1$\rightarrow $3$\rightarrow $2$\leftarrow $1$\leftarrow $3$\leftarrow $1$\rightarrow $2$\rightarrow $3$\rightarrow $1$\leftarrow $2$\leftarrow $3$\rightarrow $1$\rightarrow $2$\rightarrow $1$\leftarrow $3$\leftarrow $2$\leftarrow $1} & 6 & 3 \\ \hline
 5 & \text{1$\rightarrow $3$\rightarrow $2$\rightarrow $1$\rightarrow $3$\rightarrow $1$\rightarrow $2$\rightarrow $3$\rightarrow $1$\leftarrow $2$\leftarrow $3$\leftarrow $1$\leftarrow $3$\leftarrow $1$\leftarrow $3$\leftarrow $2$\leftarrow $1} & 6 & 1 & 23 & \text{1$\rightarrow $3$\leftarrow $2$\rightarrow $1$\rightarrow $3$\rightarrow $2$\leftarrow $1$\leftarrow $3$\leftarrow $1$\rightarrow $2$\rightarrow $3$\rightarrow $1$\rightarrow $2$\leftarrow $3$\leftarrow $1$\leftarrow $2$\leftarrow $1} & 6 & 3 \\ \hline
 6 & \text{1$\rightarrow $3$\rightarrow $1$\rightarrow $2$\rightarrow $1$\rightarrow $3$\rightarrow $1$\leftarrow $2$\leftarrow $3$\leftarrow $1$\leftarrow $3$\rightarrow $2$\rightarrow $3$\leftarrow $1$\leftarrow $3$\leftarrow $2$\leftarrow $1} & 6 & 2 & 24 & \text{1$\rightarrow $3$\rightarrow $1$\rightarrow $2$\leftarrow $3$\rightarrow $1$\rightarrow $3$\rightarrow $2$\leftarrow $1$\leftarrow $3$\leftarrow $1$\rightarrow $2$\leftarrow $3$\leftarrow $1$\leftarrow $3$\rightarrow $2$\leftarrow $1} & 6 & 4 \\ \hline
 7 & \text{1$\rightarrow $3$\rightarrow $1$\rightarrow $2$\rightarrow $3$\leftarrow $1$\rightarrow $2$\rightarrow $3$\rightarrow $1$\rightarrow $3$\leftarrow $2$\leftarrow $1$\leftarrow $3$\leftarrow $1$\leftarrow $3$\leftarrow $2$\leftarrow $1} & 6 & 2 & 25 & \text{1$\rightarrow $3$\rightarrow $1$\rightarrow $2$\leftarrow $3$\leftarrow $1$\leftarrow $3$\leftarrow $1$\rightarrow $2$\leftarrow $3$\rightarrow $1$\rightarrow $3$\rightarrow $2$\leftarrow $1$\leftarrow $3$\rightarrow $2$\leftarrow $1} & 6 & 4 \\ \hline
 8 & \text{1$\rightarrow $3$\rightarrow $1$\rightarrow $2$\rightarrow $3$\leftarrow $1$\leftarrow $3$\leftarrow $2$\leftarrow $1$\leftarrow $3$\leftarrow $1$\rightarrow $2$\rightarrow $3$\rightarrow $1$\rightarrow $3$\leftarrow $2$\leftarrow $1} & 6 & 2 & 26 & \text{1$\rightarrow $3$\rightarrow $1$\leftarrow $2$\rightarrow $3$\rightarrow $1$\leftarrow $2$\leftarrow $3$\leftarrow $1$\rightarrow $2$\rightarrow $1$\leftarrow $3$\rightarrow $2$\rightarrow $1$\leftarrow $3$\leftarrow $2$\leftarrow $1} & 6 & 4 \\ \hline
 9 & \text{1$\rightarrow $3$\rightarrow $1$\leftarrow $2$\leftarrow $3$\leftarrow $1$\leftarrow $3$\rightarrow $2$\rightarrow $1$\rightarrow $3$\rightarrow $1$\rightarrow $2$\rightarrow $3$\leftarrow $1$\leftarrow $3$\leftarrow $2$\leftarrow $1} & 6 & 2 & 27 & \text{1$\rightarrow $3$\rightarrow $1$\leftarrow $2$\rightarrow $3$\leftarrow $1$\leftarrow $3$\leftarrow $2$\rightarrow $1$\rightarrow $3$\rightarrow $1$\rightarrow $2$\leftarrow $3$\leftarrow $1$\leftarrow $3$\rightarrow $2$\leftarrow $1} & 6 & 4 \\ \hline
 10 & \text{1$\rightarrow $3$\rightarrow $1$\rightarrow $3$\rightarrow $1$\leftarrow $2$\leftarrow $3$\leftarrow $1$\leftarrow $3$\rightarrow $2$\rightarrow $1$\rightarrow $2$\rightarrow $3$\leftarrow $1$\leftarrow $3$\leftarrow $2$\leftarrow $1} & 6 & 2 & 28 & \text{1$\rightarrow $3$\rightarrow $1$\leftarrow $2$\leftarrow $3$\leftarrow $1$\rightarrow $2$\rightarrow $3$\leftarrow $1$\rightarrow $2$\rightarrow $1$\leftarrow $3$\leftarrow $2$\leftarrow $1$\rightarrow $3$\rightarrow $2$\leftarrow $1} & 6 & 4 \\ \hline
 11 & \text{1$\rightarrow $3$\rightarrow $1$\rightarrow $3$\leftarrow $2$\leftarrow $1$\leftarrow $3$\leftarrow $1$\rightarrow $2$\rightarrow $3$\rightarrow $1$\rightarrow $2$\rightarrow $3$\leftarrow $1$\leftarrow $3$\leftarrow $2$\leftarrow $1} & 6 & 2 & 29 & \text{1$\rightarrow $3$\rightarrow $1$\rightarrow $3$\rightarrow $1$\leftarrow $2$\rightarrow $3$\leftarrow $1$\leftarrow $3$\leftarrow $2$\rightarrow $1$\rightarrow $2$\leftarrow $3$\leftarrow $1$\leftarrow $3$\rightarrow $2$\leftarrow $1} & 6 & 4 \\ \hline
 12 & \text{1$\rightarrow $3$\rightarrow $2$\rightarrow $1$\leftarrow $3$\rightarrow $2$\rightarrow $1$\rightarrow $2$\rightarrow $3$\rightarrow $1$\leftarrow $2$\leftarrow $3$\leftarrow $1$\leftarrow $2$\leftarrow $3$\leftarrow $2$\leftarrow $1} & 6 & 2 & 30 & \text{1$\rightarrow $3$\rightarrow $1$\rightarrow $3$\rightarrow $2$\leftarrow $1$\leftarrow $3$\leftarrow $1$\rightarrow $2$\leftarrow $3$\rightarrow $1$\rightarrow $2$\leftarrow $3$\leftarrow $1$\leftarrow $3$\rightarrow $2$\leftarrow $1} & 6 & 4 \\ \hline
 13 & \text{1$\rightarrow $3$\leftarrow $2$\leftarrow $1$\leftarrow $3$\rightarrow $2$\rightarrow $1$\rightarrow $2$\rightarrow $3$\rightarrow $1$\rightarrow $2$\rightarrow $3$\leftarrow $1$\leftarrow $2$\leftarrow $3$\leftarrow $2$\leftarrow $1} & 6 & 2 & 31 & \text{1$\rightarrow $3$\rightarrow $1$\rightarrow $2$\rightarrow $3$\leftarrow $1$\rightarrow $2$\leftarrow $3$\rightarrow $1$\leftarrow $2$\leftarrow $1$\leftarrow $3$\leftarrow $2$\rightarrow $1$\leftarrow $3$\rightarrow $2$\leftarrow $1} & 6 & 5 \\ \hline
 14 & \text{1$\rightarrow $3$\rightarrow $1$\rightarrow $2$\rightarrow $1$\rightarrow $3$\rightarrow $1$\leftarrow $2$\rightarrow $3$\leftarrow $1$\leftarrow $3$\leftarrow $2$\leftarrow $3$\leftarrow $1$\leftarrow $3$\rightarrow $2$\leftarrow $1} & 6 & 3 & 32 & \text{1$\rightarrow $3$\rightarrow $1$\rightarrow $3$\rightarrow $1$\leftarrow $2$\rightarrow $3$\leftarrow $1$\rightarrow $2$\leftarrow $3$\leftarrow $1$\leftarrow $3$\leftarrow $2$\rightarrow $1$\leftarrow $3$\rightarrow $2$\leftarrow $1} & 6 & 5 \\ \hline
 15 & \text{1$\rightarrow $3$\rightarrow $1$\rightarrow $2$\leftarrow $3$\leftarrow $1$\rightarrow $2$\rightarrow $3$\rightarrow $1$\leftarrow $2$\leftarrow $1$\leftarrow $3$\rightarrow $2$\rightarrow $1$\leftarrow $3$\leftarrow $2$\leftarrow $1} & 6 & 3 & 33 & \text{1$\rightarrow $3$\leftarrow $2$\leftarrow $1$\leftarrow $3$\leftarrow $1$\rightarrow $2$\leftarrow $3$\rightarrow $1$\leftarrow $2$\rightarrow $3$\rightarrow $1$\rightarrow $2$\rightarrow $1$\leftarrow $3$\rightarrow $2$\leftarrow $1} & 6 & 5 \\ \hline
 16 & \text{1$\rightarrow $3$\rightarrow $1$\rightarrow $2$\leftarrow $3$\leftarrow $1$\leftarrow $2$\rightarrow $3$\leftarrow $1$\rightarrow $2$\rightarrow $1$\rightarrow $3$\rightarrow $2$\leftarrow $1$\leftarrow $3$\leftarrow $2$\leftarrow $1} & 6 & 3 & 34 & \text{1$\rightarrow $3$\rightarrow $1$\leftarrow $2$\rightarrow $3$\leftarrow $1$\rightarrow $2$\leftarrow $3$\leftarrow $1$\rightarrow $2$\rightarrow $1$\leftarrow $3$\rightarrow $2$\leftarrow $1$\rightarrow $3$\leftarrow $2$\leftarrow $1} & 6 & 6 \\ \hline
 17 & \text{1$\rightarrow $3$\rightarrow $1$\rightarrow $2$\leftarrow $3$\leftarrow $1$\leftarrow $3$\leftarrow $1$\leftarrow $3$\leftarrow $2$\rightarrow $1$\rightarrow $3$\rightarrow $1$\leftarrow $2$\rightarrow $3$\rightarrow $2$\leftarrow $1} & 6 & 3 & 35 & \text{1$\rightarrow $3$\rightarrow $1$\leftarrow $2$\leftarrow $3$\rightarrow $1$\leftarrow $2$\rightarrow $3$\leftarrow $1$\rightarrow $2$\rightarrow $1$\leftarrow $3$\leftarrow $2$\rightarrow $1$\leftarrow $3$\rightarrow $2$\leftarrow $1} & 6 & 6 \\ \hline
 18 & \text{1$\rightarrow $3$\rightarrow $1$\rightarrow $2$\leftarrow $3$\leftarrow $1$\leftarrow $3$\rightarrow $2$\rightarrow $1$\rightarrow $3$\rightarrow $1$\leftarrow $2$\rightarrow $3$\leftarrow $1$\leftarrow $3$\leftarrow $2$\leftarrow $1} & 6 & 3 &  &  &  &  \\ \hline
\hline
\end{tabular}
}
\end{center}
\caption{16-link TDI combinations that suppress laser noise. $\text{N}_s$ is the sequence number. Numbers within each sequence indicate the satellite on which each event takes place. First and last event coincide. Arrow indicate the direction of the link connecting adjoining events. Events at the start of two arrows represent simultaneous emission of two beams. Events at the end of two arrows represent measurements. The first  is always an emission event. $\text{N}_l$ and $\text{N}_m$ are the number of inter-satellite links and the number of measurement events involved in the sequence respectively.}
\label{T16l}
\end{table}

\section{Discussion}
Our 174 16-link, noise suppressing TDI combination are definitely more numerous than the 48  combinations counted by ref. \cite{Vallisneri} . In addition, while we  find 12 12-link combinations,  ref \cite{Vallisneri} found none. 

The difference is in  the search criterion.  The search of  ref \cite{Vallisneri}   is based on eq. \ref{Edl} that leaves completely open the effect of the interchange of the two satellites at the end of a link:
\begin{equation}
\begin{split}
L_{ij} \ne L_{ji}\\ 
\dot{L}_{ij} \ne \dot{L}_{ji},
\label{ecel}
\end{split}
\end{equation}
If we apply this same non-symmetric  criterion to our 174 16-link combinations, indeed only 48 remain. We have not performed yet a detailed analysis to clarify if these 48 combinations coincide with those found in   \cite{Vallisneri}, however  the basic  X, Y and Z combinations certainly belong to both sets. In addition, none of our 12-link combinations survive the test.  

\begin{figure}[h]
\includegraphics[width=0.6\textwidth]{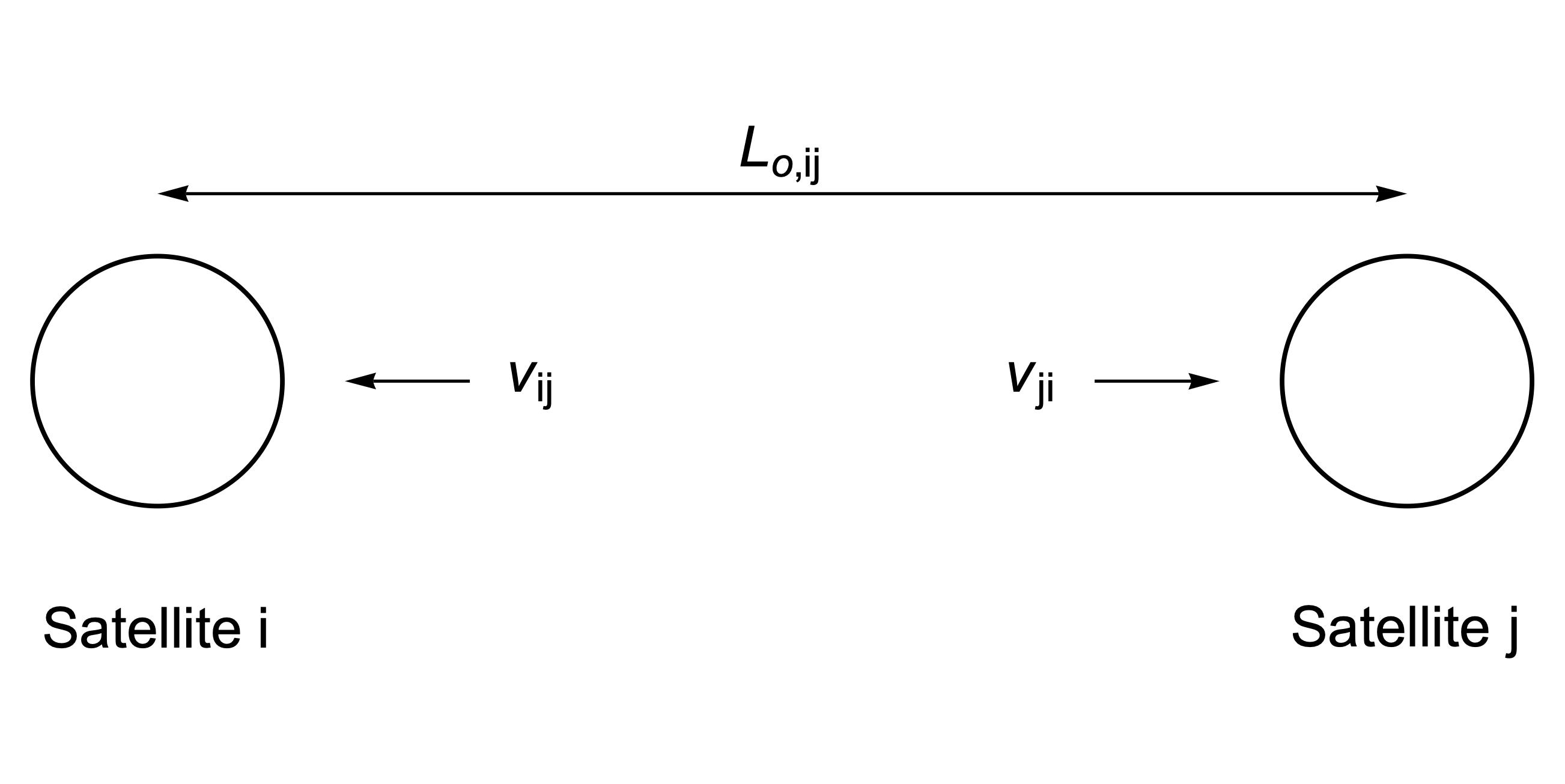}
\caption{Schematics of the velocity convention}
\label{Figv}
\end{figure}

The formulas  for $L_{ij}$ and $\dot{L}_{ij}$ in eq. \ref{sym} that we have used for our search,  show instead some relevant  symmetries. Let us define $v_{ij}$ and $v_{ji}$ as the components of the velocities of emitter and receiver respectively, both taken  along the direction of $\Delta \hat{r}$, {\it but each taken positive when pointing away from the other satellite}, see fig. \ref{Figv}. 

Let's also define $L_{o,ij}=\vert\Delta \vec{r}_{ij}\vert$. Then:
\begin{equation}
\begin{split}
L_{ij}=L_{o,ij}\left(1-\frac{v_{ij}}{c}\right)\\
L_{ji}=L_{o,ij}\left(1-\frac{v_{ji}}{c}\right)\\ 
\dot{L}_{ij} = \dot{L}_{ji}=v_{ij}+v_{ji}\\ 
\label{ecom}
\end{split}
\end{equation}

These differences have significant consequences. For instance, for the sequence of fig. \ref{Figseq},  assuming that only the more general eq. \ref{ecel} holds, we get:
\begin{equation}
\begin{split} 
&t_{13}= t_1+\\
&+\frac{1}{c^2}\left(\left(L_{13}+L_{21}-L_{23}\right) \left(\dot{L}_{12}+\dot{L}_{31}-\dot{L}_{32}\right)-\left(L_{12}+L_{31}-L_{32}\right) \left(\dot{L}_{13}+\dot{L}_{21}-\dot{L}_{23}\right)\right)
\end{split}
\end{equation}
which is first order in $\dot{L}/c$. If instead the more restrictive eq. \ref{ecom} holds, the quantity above is found to be second order in $v/c$, and then negligible, and then $t_{13}=t_1$. 

Same is true for any other of the 12 link combinations we have found, for which eq. \ref{ecel} gives always $t_{13}\ne t_1$.
In particular for  $\alpha$, as already noted in  \cite{Shaddock2003}, we get: 
\begin{equation}
\begin{split} 
& t_{13}= t_1+\\
&+\frac{1}{c^2}\left(\left(L_{13}+L_{32}+L_{21}\right) \left(\dot{L}_{12}+\dot{L}_{23}+\dot{L}_{31}\right)-\left(L_{12}+L_{23}+L_{31}\right) \left(\dot{L}_{13}+\dot{L}_{32}+\dot{L}_{21}\right)\right)
\end{split}
\label{eldn}
\end{equation}
while, again, in the quasi-inertial frame we get $t_{13}=t_1$.
 
With the aim of resolving this apparent discrepancy, we have evaluated  $\delta t\equiv t_{13}-t_1$ in eq. \ref{eldn}, \emph{in the heliocentric frame}, in the case of $\alpha$. We have used eq. \ref{Elldot} to estimate the various quantities in eq. \ref{eldn},  and found that, to the leading order:
 \begin{equation}
\begin{split} 
&\delta t\simeq  \frac{3 L}{c} \left(\frac{\left(\vec{a}_1+\vec{a}_3\right)\cdot\Delta \vec{r}_{13}}{c^2}+\frac{\left(\vec{a}_2+\vec{a}_3\right)\cdot\Delta \vec{r}_{32}}{c^2}+\frac{\left(\vec{a}_2+\vec{a}_1\right)\cdot\Delta \vec{r}_{21}}{c^2}\right)
\end{split}
\label{lde}
\end{equation}
with $L$ the average length of the three arms, and  $\vec{a}_i$  the acceleration  of the $\text{i}^{\text{th}}$ satellite.

We have calculated numerically the right hand side of eq. \ref{lde} for the LISA orbits. We find that, despite the individual terms  are of order of nano-seconds,  they compensate each other to a large extent, so that $\delta t$, which oscillates with one year period, remains  limited to $\vert\delta t\vert\le 3\times10^{-13} s$. This is  smaller than or, at most, comparable to  the numerical accuracy of our numerical results (see table \ref{T:X}).

To better understand the origin of $\delta t$, and the reasons for  such  suppression, we have  also calculated the theoretical value of the right hand side of eq. \ref{lde}, in the approximation where  the   LISA motion  results from  the superposition of the rigid body motion of an equilateral triangle, with the in-plane distortion of the triangle itself (see appendix sect. \ref{A2} for definitions and for the calculation details). We find that, to first order in the amplitude of the distortion, $\delta t$ is given by:   
\begin{equation}
\begin{split} 
 &\delta t = \frac{3L}{c}\left(\frac{4\frac{d \Omega_n}{dt}A+4 \Omega_n\frac{dA}{dt}}{c^2}\right)= \frac{3L}{c}\frac{d}{dt}\left(\frac{4\Omega_nA}{c^2}\right)
\end{split}
\label{app}
\end{equation}
where $\Omega_n$ is the component of the angular velocity normal to the triangle and $A$ is the triangle area . In the appendix sect \ref{A2} we derive a rough numerical estimate for this quantity  and find $\delta t\simeq\frac{3L}{c}\frac{d}{dt}\left(\frac{4\Omega_nA}{c^2}\right)\le5 \times10^{-13} s$, in reasonable agreement with the figure  calculated numerically from the orbits.

Thus $\delta t$ is equal to the total light travel time around the constellation, multiplied by the time derivative of the notorious Sagnac formula $4\Omega_nA$ for the difference of light travel time around a rotating closed loop.
By inspecting table \ref{T:X}, one can note that $\alpha$ may indeed be seen  as the difference between two Sagnac interferometers, delayed, one relative to the other,  by the light round trip time $3L/c$ . Thus the formula in eq. \ref{app} gives approximately the change in the difference of the round trip time between these two delayed Sagnac interferometers. 

Note that this quantity is independent of the reference frame, heliocentric or quasi-inertial, as these two systems have no relative rotation. Thus, in the heliocentric frame, the right hand side of eq. \ref{eldn} is only apparently first order in $\dot{L}/c$, but is in reality fully negligible. We  think that this illustrates clearly enough  how the extra corrections to the delays, which appear in the heliocentric frame, in reality cancel each other out due to the nearly rigid motion motion of the constellation.

Finally, it is important to note  that actual suppression of phase noise in LISA will depend on many other technical details. Performance of satellite-to-satellite ranging,  of the frequency distribution system, and of the sophisticated sequence of filtering, sampling and interpolation of actual data, will set the final figure \cite{Otto_2012,filters}. 
Nevertheless the results presented here open up the possibility to study if  these additional TDI combinations, once implemented in the actual measurement scheme, may be used  for instrument characterisation, both for noise discrimination and for cancellation of non astrophysical signals. We are currently running such a study and we will report its results in a future paper.

\newpage
\section{\label{A1}Appendix 1. Detailed calculation of delay}

To calculate the delay, we approximate the  positions of the emitter at the time of the emission $t_e=t-\Delta t$,  and of the receiver at the time of reception $t_r=t$, as:
\begin{equation}
\begin{split}
&\vec{r}_e\left(t-\Delta t\right)=\vec{r}_o+\vec{v}  \left(t-\Delta t\right)+\frac{1}{2} \vec{a} \left(t-\Delta t\right)^2\\
&\vec{r}_r\left(t\right)=\vec{r}_o+\Delta \vec{r}+\left(\vec{v}+\Delta \vec{v}\right) t+\frac{1}{2}\left(\vec{a}+\Delta \vec{a}\right)t^2
\end{split}
\label{Epos2}
\end{equation}
Here we have made the  assumption  that  the satellites are moving with constant accelerations, at least during the total time of propagation of  light throughout the constellation.  This approximation is reasonably good: with the nominal LISA orbits from ref. \cite{Hughes}, the difference between the exact orbit and the constant acceleration approximation is less than a millimetre over an elapsed time of 200 seconds. Such a  time lapse is definitely longer than the total time of light propagation  in any of the TDI combinations of interest here. 

We can plug the expressions in eq. \ref{Epos2} into the condition in eq. \ref{Enull} , and  expand the result with Mathematica to get:
\begin{equation}
\begin{split}
&\Delta r^2+2  \Delta \vec{v}\cdot\Delta \vec{r}\; t+\left(\Delta \vec{a}\cdot\Delta \vec{r}+\Delta v^2\right)t^2+\Delta \vec{a}\cdot \Delta \vec{v} \;t^3+\frac{1}{4}\Delta a^2 t^4+\\ 
&+\Delta t\left( 2 \left(\vec{v}\cdot\Delta \vec{r}-c \delta r\right)+2\;\left(\vec{a}\cdot\Delta \vec{r}+\vec{v}\cdot\Delta \vec{v}\right)t+\left(2\;\vec{a}\cdot\Delta \vec{v}+\vec{v}\cdot\Delta \vec{a}\right)t^2+\vec{a}\cdot\Delta \vec{a}\;t^3\right)+\\
&+\Delta t^2\left(-c^2+v^2-\vec{a}\cdot\Delta \vec{r}+\left(2\vec{a}\cdot \vec{v}-\vec{a}\cdot\Delta \vec{v}\right)t+\left(a^2-\frac{\vec{a}\cdot\Delta \vec{a}}{2}\right)t^2\right)+\\
&-\Delta t^3\left(\vec{a}\cdot \vec{v}+a^2 t\right)+\Delta t^4 \frac{a^2}{4}=0,
\label{Enulexp}
\end{split}
\end{equation}
where $\delta r/c$ is the Shapiro delay  which, within the requested accuracy,  does not vary appreciably over the propagation time associated with a TDI combination.

Let's now define $u=c t/\vert \Delta \vec{r}\vert$, $\Delta u=c \Delta t/\vert \Delta \vec{r}\vert$, $v_a^2=\vert \vec{a}\vert \vert \Delta \vec{r}\vert$, and finally $\Delta v_a^2=\vert \Delta \vec{a}\vert \vert \Delta \vec{r}\vert$. In addition for any vector $\vec{x}$, we call $\hat{x}$  the unit vector parallel to it. Eq. \ref{Enulexp} becomes then:
\begin{equation}
\begin{split}
&\left(1+2 \frac{ \Delta \vec{v}}{c}\cdot\Delta \hat{r}\; u+\left(\frac{\Delta v_a^2}{c^2}\Delta \hat{a}\cdot\Delta \hat{r}+\frac{\Delta v^2}{c^2}\right)u^2 +\frac{\Delta v_a^2}{c^2}\Delta \hat{a}\cdot \frac{\Delta \vec{v}}{c} \;u^3 +\frac{1}{4}\frac{\Delta v_a^4}{c^4} u^4 \right)+\\
&+\Delta u \left(2\left(\frac{\vec{v}}{c}\cdot\Delta \hat{r}-\frac{\delta r}{\Delta r}\right)+2\;\left(\frac{v_a^2}{c^2}\hat{a}\cdot\Delta \hat{r}+\frac{\vec{v}}{c}\cdot \frac{\Delta\vec{v}}{c}\right)u\right.\\
&\left.+\left(2\;\frac{v_a^2}{c^2}\hat{a}\cdot \frac{\Delta\vec{v}}{c}+\frac{\Delta v_a^2}{c^2}\frac{\vec{v}}{c}\cdot\Delta \hat{a}\right)u^2 +\frac{v_a^2 }{c^2}\frac{\Delta v_a^2}{c^2}\hat{a}\cdot\Delta \hat{a}\;u^3 \right)+\\ 
&+\Delta u^2 \left(-1+\frac{v^2}{c^2}-\frac{v_a^2}{c^2}\hat{a}\cdot\Delta \hat{r}+\frac{v_a^2}{c^2}\left(2\hat{a}\cdot \frac{\vec{v}}{c}-\hat{a}\cdot\frac{\Delta \vec{v}}{c}\right)u+\left(\frac{v_a^4}{c^4}-\frac{v_a^2}{c^2}\frac{ \Delta v_a^2}{c^2}\frac{\hat{a}\cdot\Delta\hat{a}}{2}\right)u^2\right)+\\
&-\Delta u^3 \left(\frac{v_a^2}{c^2}\hat{a}\cdot \frac{\vec{v}}{c}+\frac{v_a^4}{c^4} u\right)+\Delta u^4 \frac{v_a^4}{4c^4}=0
\label{Enulexp2}
\end{split}
\end{equation}

The various  terms in eq. \ref{Enulexp}, have rather different values depending on the reference frame one uses.

We begin with the  celestial isotropic frame. Quantities may be estimated  using nominal LISA orbits as specified in \cite{Hughes}. This choice however  ignores the gravitational perturbations of the other bodies. To be on the safe side we have  also checked the results with  the orbits resulting from ESA's Concurrent Design Facility study run in 2017 \footnote{ESA CDF team, private communication} obtaining comparable results.
\begin{enumerate}
\item{$\Delta u\simeq 1$ and $u$ may be at most $\simeq 8$. We need to know the delay within the equivalent of 1 light-meter. With a $2.5 \;Gm$ arm-length, this correspond to a required relative resolution on $\Delta u$ of $\simeq 4\times10^{-10}$}
\item{$\vert\Delta v\vert \lesssim 1\;m/s$ for the nominal orbit, and $\vert\Delta v\vert \lesssim 10\;m/s$ for the simulated orbit. this gives at most  $\Delta v /c\simeq 3\times10^{-8}$ so that $\left(\Delta v /c\right)^2\simeq 10^{-15}$.}
\item{$v/c\simeq \sqrt{\frac{R_\odot}{2R}}\simeq 10^{-4}$. Here $R$ is one astronomical unit and $R_\odot$ is the Schwarzschild radius of the Sun. This rough estimate is confirmed with simulated orbits. Thus $\left(v/c\right)^3> \left(\Delta v /c\right)^2$.}
\item{$\frac{v_a^2}{c^2}\simeq \frac{R_\odot}{2R}\frac{\vert \Delta \vec{r}\vert}{R}\simeq 2\times 10^{-10}$. Simulated orbits give that indeed $\frac{\vert \Delta r \cdot \vec{a}\vert}{c^2}\lesssim9\times10^{-11}$, the small reduction being probably due to the angular factor.}
\item{For these low eccentricity orbits the differential acceleration is $\Delta \vec{a}\simeq \frac{R_\odot}{2R^3}\Delta \vec{r}$. This gives $\frac{\Delta v_a^2}{c^2}\simeq \frac{R_\odot}{2R}\left(\frac{\vert \Delta \vec{r}\vert}{R}\right)^2\simeq 3\times 10^{-12}$. The simulated orbits confirm such estimate to the second digit.}
\item{$\frac{\delta r}{\Delta r}\simeq 2\times 10^{-8}$ which justifies the choice to take only first order terms in the Shapiro delay}
\end{enumerate}
We proceed by initially  keeping all terms  larger than $\left(\Delta v /c\right)^2$. Eq. \ref{Enulexp} simplifies then to:
\begin{equation}
\begin{split}
&\left(1+2 \frac{ \Delta \vec{v}}{c}\cdot\Delta \hat{r}\; u+\frac{\Delta v_a^2}{c^2}\Delta \hat{a}\cdot\Delta \hat{r}\;u^2 \right)+\\
&+\Delta u \left(2\left(\frac{\vec{v}}{c}\cdot\Delta \hat{r}-\frac{\delta r}{\Delta r}\right)+2\;\left(\frac{v_a^2}{c^2}\hat{a}\cdot\Delta \hat{r}+\frac{\vec{v}}{c}\cdot \frac{\Delta\vec{v}}{c}\right)u\right)+\Delta u^2 \left(-1+\frac{v^2}{c^2}-\frac{v_a^2}{c^2}\hat{a}\cdot\Delta \hat{r}\right)=0,
\label{Esimp}
\end{split}
\end{equation}
the physically meaningful solution of which, expanded to same order, is:
\begin{equation}
\begin{split}
&\Delta u=1+\frac{\vec{v}}{c}\cdot \Delta \hat{r}-\frac{\delta r}{\Delta r}+\frac{1}{2}\left(\frac{\vec{v}}{c}\cdot \Delta\hat{r}\right)^2+\frac{1}{2}\frac{v^2}{c^2}\left(1+2\;\frac{\vec{v}}{c}\cdot \Delta \hat{r}\right)-\frac{1}{2}\frac{v_a^2}{c^2}\hat{a}\cdot\Delta \hat{r}+\\
&+ u\left(\frac{v_a^2}{c^2}\hat{a}\cdot\Delta \hat{r}+\frac{\Delta\vec{v}}{c}\cdot\Delta\hat{r}+\frac{\Delta \vec{v}}{c}\cdot \frac{\vec{v}}{c}\right)+\frac{1}{2}u^2\frac{ \Delta v_a^2}{c^2}\Delta \hat{a}\cdot\Delta\hat{r}
\label{Esol}
\end{split}
\end{equation}
We now recover the more relaxed requirement of  a relative uncertainty of $4\times 10^{-10}$. If one assumes $u\le10$,  neglects terms that are less than $10^{-10}$, and restores the meaning of some of the symbols, he gets the final result for the heliocentric frame:
\begin{equation}
\begin{split}
&\Delta t\left(t\right)=\frac{\vert \Delta \vec{r}\vert}{c}\times\left(1+\frac{\vec{v}}{c}\cdot \Delta\hat{r}-\frac{\delta r}{\Delta r}+\frac{1}{2}\left(\frac{\vec{v}}{c}\cdot \Delta\hat{r}\right)^2+\frac{1}{2}\frac{v^2}{c^2}-\frac{1}{2}\frac{\vec{a}\cdot\Delta \vec{r}}{c^2}\right)+\\
&+ t\;\left(\frac{\vec{a}\cdot\Delta \vec{r}}{c^2}+\frac{\Delta\vec{v}}{c}\cdot\Delta\hat{r}\right)
\label{Efin}
\end{split}
\end{equation}
which, in the language of effective arm-length and arm-length time derivative, is equivalent to:
\begin{equation}
\begin{split}
&L=\vert \Delta \vec{r}\vert\times\left(1+\frac{\vec{v}}{c}\cdot \Delta\hat{r}-\frac{\delta r}{\Delta r}+\frac{1}{2}\left(\frac{\vec{v}}{c}\cdot \Delta\hat{r}\right)^2+\frac{1}{2}\frac{v^2}{c^2}-\frac{1}{2}\frac{\vec{a}\cdot\Delta \vec{r}}{c^2}\right)\\
&\\
&\dot{L}=\left(\frac{\vec{a}\cdot\Delta \vec{r}}{c}+\Delta\vec{v}\cdot\Delta\hat{r}\right)
\end{split}
\label{Elldot}
\end{equation}

For the quasi-inertial frame instead
\begin{enumerate}
\item{$v/c\le 3\times10^{-7}$.  Thus $\left(v/c\right)^2 \ll 10^{-10}$.}
\item{$\frac{v_a^2}{c^2}  < 2\times 10^{-12}\ll 10^{-10}$.}
\item{ $\frac{\Delta v_a^2}{c^2}\simeq 3\times 10^{-12} \ll 10^{-10}$.}
\item{$\frac{\delta r}{\Delta r}\simeq 2\times 10^{-12}$}
\end{enumerate}

Eq. \ref{Enulexp} simplifies then to:
\begin{equation}
\begin{split}
&\left(1+2 \frac{ \Delta \vec{v}}{c}\cdot\Delta \hat{r}\; u \right)+\Delta u \left(2  \frac{\vec{v}}{c}\cdot\Delta \hat{r}\right)-\Delta u^2 =0,
\label{Esimp}
\end{split}
\end{equation}
the physically meaningful solution of which, expanded to same order, is:
\begin{equation}
\begin{split}
&\Delta u=1+\frac{\vec{v}}{c}\cdot \Delta \hat{r}+ \frac{\Delta\vec{v}}{c}\cdot\Delta\hat{r}\; u
\label{Esol}
\end{split}
\end{equation}
Restoring the meaning of some of the symbols, we get:
\begin{equation}
\begin{split}
&\Delta t\left(t\right)=\frac{\vert \Delta \vec{r}\vert}{c}\times\left(1+\frac{\vec{v}}{c}\cdot \Delta\hat{r}\right)+\frac{\Delta\vec{v}}{c}\cdot\Delta\hat{r}\; t
\label{Efin}
\end{split}
\end{equation}.
\section{\label{A2} Appendix 2. Calculation of $\delta t$}
\begin{figure}[h]
\includegraphics[width=0.9\textwidth]{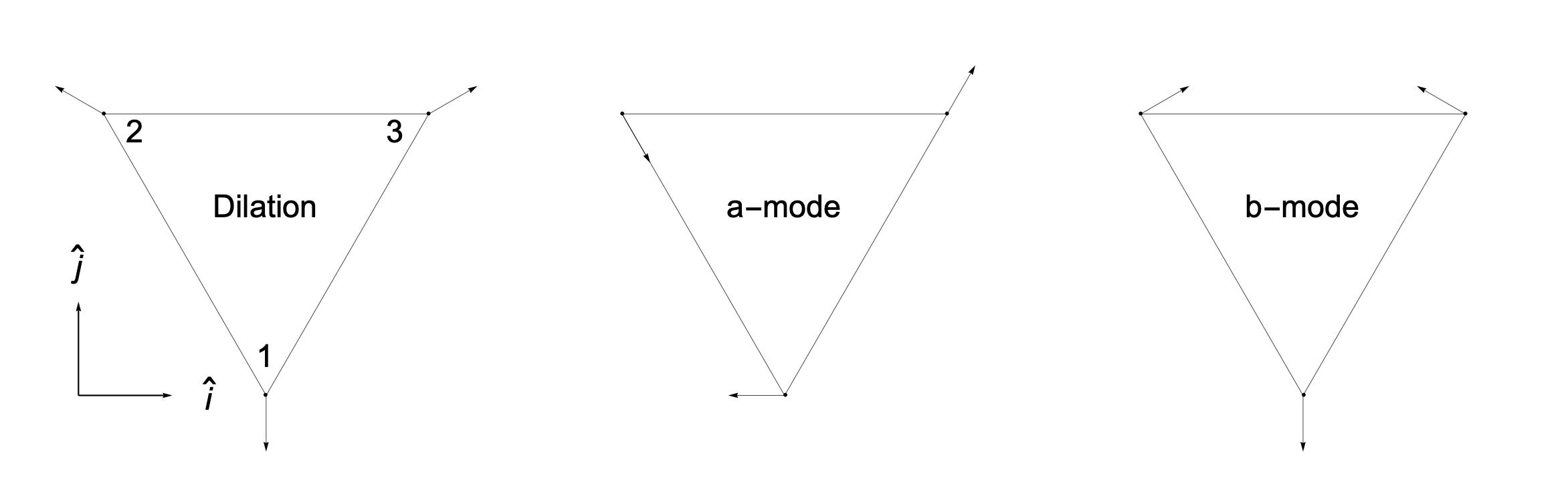}
\caption{Distortion modes of a triangle. Also shown are the unit vector of the x-axis $\hat{i}$,  that of the y-axis $\hat{j}$, and the numbering of satellites.}
\label{Figmodes}
\end{figure}

As the three satellites always define a plane, we approximate the motion of the constellation as the superposition of the rigid rotation of its plane, superimposed to the in-plane distortion of a nominally equilateral triangle. The motion of  a triangle can notoriously be analysed in terms of six normal modes  \cite{normalmodes}. Of these six modes, three correspond to in-plane rigid rotation and translation, and are then, in our case, adsorbed into the rigid motion of the plane, and three correspond to in-plane distortion of the triangle and are illustrated in fig. \ref{Figmodes}.

The  unit vectors for the motion of the satellites are given in table \ref{tmodes}.
\begin{table}[h]
\begin{tabular}{|c|c|c|c|}
\hline
&Satellite 1&Satellite 2&Satellite 3\\
\hline
Dilation&$-\hat{j}$&$\frac{\hat{j}}{2}-\frac{\sqrt{3} \hat{i}}{2}$&$\frac{\sqrt{3} \hat{i}}{2}+\frac{\hat{j}}{2}$\\
\hline
a-mode&$-\hat{i}$&$\frac{\hat{i}}{2}-\frac{\sqrt{3} \hat{j}}{2}$&$\frac{\hat{i}}{2}+\frac{\sqrt{3} \hat{j}}{2}$\\
\hline
b-mode&$-\hat{j}$&$ \frac{\sqrt{3} \hat{i}}{2}+\frac{\hat{j}}{2} $&$ \frac{\hat{j}}{2}-\frac{\sqrt{3} \hat{i}}{2}$\\
\hline
\end{tabular}
\caption{Unit vectors  for the motion of satellites for the three distortion modes.}
\label{tmodes}
\end{table}
Within such an approximation, the acceleration of satellite $i$ is given by
\begin{equation}
\begin{split}
\vec{a}_i=\vec{a}_c+\frac{d\vec{\Omega}}{dt}\times \vec{r'}_i+2 \vec{\Omega}\times \vec{v'}_i+\vec{\Omega}\times\left(\vec{\Omega}\times \vec{r'}_i\right)+\vec{a'}_i
\end{split}
\label{eacc}
\end{equation}
Here $\vec{a}_c$ and $\vec{\Omega}$ are, respectively,  the acceleration of the  constellation center and the  angular velocity of the plane of the constellation, both relative to the heliocentric frame. Vectors $\vec{r'}_i$, $\vec{v'}_i$ and  $\vec{a'}_i$ are the in-plane satellite position, velocity and  acceleration respectively, relative to the constellation center.

One can use eq. \ref{eacc} to calculate the various terms in eq. \ref{lde}. The calculation follows. Note  that in all equations below we use the offset modulo sum defined by $i+1= 2,3,1$ for $i=1,2,3$ respectively.
\begin{myitemize}
\item{One can easily recognise that, for a triangle
\begin{equation}
\sum_{i=1}^3\vec{a}_c\cdot\Delta\vec{r'}_{i+1,i}=\vec{a}_c\cdot\sum_{i=1}^3\Delta\vec{r'}_{i+1,i}=0
\end{equation}
Here we have used the notion that $\Delta \vec{r}_{i,j}=\Delta \vec{r'}_{i,j} $}
\item{Also, with a little patience one gets that
\begin{equation}
\begin{split}
&\sum_{i=1}^3\vec{\Omega}\times\left(\vec{\Omega}\times\left( \vec{r'}_{i+1}+ \vec{r'}_i\right)\right)\cdot \Delta\vec{r'}_{i+1, i}=\\
&\sum_{i=1}^3\left(\vec{\Omega}\cdot \vec{r'}_{i+1}\right)^2-\sum_{i=1}^3\left(\vec{\Omega}\cdot \vec{r'}_{i}\right)^2-\Omega^2 \sum_{i=1}^3\left(r_{i+1}^{'2}-r_{i}^{'2}\right)=0
\end{split}
\end{equation}
}
\item{Calling $\hat{u}_{ki}$ the unit vector of the motion of satellite $k$ for mode $i$ (given in table \ref{tmodes})  one can check that, for all modes
\begin{equation}
\sum_{i=1}^3\left(\hat{u}_{k,i+1}+\hat{u}_{k,i}\right)\cdot\left(\vec{r'}_{i+1}-\vec{r'}_{i}\right)=0
\label{emodes}
\end{equation}
thus $\vec{a'}$ does not contribute to $\delta t$.}
\item{The contribution of angular acceleration gives
\begin{equation}
\begin{split}
&\sum_{i=1}^3\frac{d\vec{\Omega}}{dt}\times \left(\vec{r'}_{i+1}+\vec{r'}_{i}\right)\cdot\left(\vec{r'}_{i+1}-\vec{r'}_{i}\right)=\frac{d\vec{\Omega}}{dt}\cdot\sum_{i=1}^3 \left(\vec{r'}_{i+1}+\vec{r'}_{i}\right)\times\left(\vec{r'}_{i+1}-\vec{r'}_{i}\right)=\\
&=2\frac{d\vec{\Omega}}{dt}\cdot\sum_{i=1}^3 \vec{r'}_{i}\times\vec{r'}_{i+1}=4\frac{d\Omega_n}{dt} A
\end{split}
\end{equation}
Here  $\Omega_n$ is the component of the angular velocity along $\vec{r'}_{i}\times\vec{r'}_{i+1}$, and then normal to the plane, and A is the triangle area. To get the final result we have used the fact that the modulus of $\vec{r'}_{i}\times\vec{r'}_{i+1}$ is twice the area of the triangle whose vertexes are the center of the constellation and satellites $i$ and $i+1$.}
\item{Finally the Coriolis term, for the $\text{k}^\text{th}$ mode, with velocity amplitude $v_o$, gives:
\begin{equation}
\begin{split}
2 v_o \sum_{i=1}^3\vec{\Omega}\times \left(\hat{u}_{k,i+1}+\hat{u}_{k,i}\right)\cdot\left(\vec{r'}_{i+1}-\vec{r'}_{i}\right)=2 v_o\vec{\Omega}\cdot \sum_{i=1}^3 \left(\hat{u}_{k,i+1}+\hat{u}_{k,i}\right)\times\left(\vec{r'}_{i+1}-\vec{r'}_{i}\right).
\end{split}
\end{equation}
Here, to first order, the $\vec{r'}_{i}$ can be taken for an undistorted equilateral triangle. With this assumption, the sum is zero for the a- and b- modes while for dilation its value is a vector normal to the plane of the triangle and with modulus 3 time the length of the triangle side.
Thus in total this term is equal to $6 L v_o \Omega_n$. This results can be put in a more interesting form by noting that, in pure dilation, the triangle remain equilateral, and that $v_o$ is the rate of change of the length of its apothem. This change of apothem induces a corresponding  change of area at a rate $\frac{dA}{dt}=\frac{3}{2}Lv_o$. By substituting in the result above,  the Coriolis term becomes $4\Omega_n \frac{dA}{dt}$.
}
\end{myitemize}
Adding up all the terms above, and substituting in eq. \ref{lde} we get indeed the results of eq \ref{app}:
\begin{equation}
\begin{split} 
 &\delta t = \frac{3L}{c}\left(\frac{4\frac{d \Omega_n}{dt}A+4 \Omega_n\frac{dA}{dt}}{c^2}\right)= \frac{3L}{c}\frac{d}{dt}\left(\frac{4\Omega_nA}{c^2}\right)
\end{split}
\label{app2}
\end{equation}

A numerical estimate of $\frac{3L}{c}\frac{d}{dt}\left(\frac{4\Omega_nA}{c^2}\right)$ can be obtained as follows. A rough upper bound to the constellation  angular acceleration may be estimated from the variation of the angular velocity of each satellite due to the eccentricity $e$ of the orbit $\Delta \Omega/\Omega\simeq 2 e$. For a sinusoidal yearly oscillation of the angular velocity, this would give a peak angular acceleration of $\dot{\Omega}\simeq  2 e \Omega^2$, with $\Omega=2\pi/(1 \text{year})$. However, the angular velocity of the LISA constellation results approximately from the superposition of a yearly rotation around an axis normal to the ecliptic, and of  another yearly rotation, with opposite sign, around the normal to the triangle. Thus the projection of the entire  angular velocity normal to the triangle is, in magnitude, just one half that around the normal to the ecliptic. With this correction, $\dot{\Omega}\simeq   e \Omega^2\simeq 2\times10^{-16} s^{-2}$.  Using this figure, $A/c^2\simeq 30 s^2$, and $3L/c\simeq 25 s$,  the term $\frac{3L}{c}\frac{4A}{c^2}\frac{d \Omega_n}{dt}$ would  peak at  $5 \times10^{-13} s$. 

As for the  term $\frac{4 \Omega_n\frac{dA}{dt}}{c^2}$ we can only estimate numerically $\frac{\delta A}{c^2}$ from the orbits. We find that, despite the  well known, comparatively large oscillation of the lengths of the three arms, $\simeq 1.2\times 10^4 km$ for $2.5\times10^6 km$ arms, the area oscillates only by  $\frac{\delta A}{c^2}\le 10^{-2}s^2$, though with a period of a third of a year, with  a peak value of the derivative at $\frac{1}{c^2}dA/dt\le 6 ns$. Using the same figures as above for delay and angular velocity, we get  $ \frac{3L}{c}\frac{4 \Omega_n\frac{dA}{dt}}{c^2}\le 6\times10^{-14} s$, which contributes negligibly to the overall correction that remains then $\delta t\simeq\frac{3L}{c}\frac{d}{dt}\left(\frac{4\Omega_nA}{c^2}\right)\le5 \times10^{-13} s$. 
\section{Acknowledgements}
We would like to thank  Jean-Baptiste Bayle, Olaf Hartwig, Luciano Iess, Antoine Petiteau,  Massimo Tinto, Michele Vallisneri,  and  the whole Trento LISA group for very insightful discussions. 
This work was supported in part by Agenzia Spaziale Italiana and Istituto Nazionale di Fisica Nucleare.
\bibliography{generic}
\end{document}